# AI-assisted search for self-compressing gigawatt subcycle field waveforms

**ALEKSEI M. ZHELTIKOV,**[1,*] **YOUCHAN PARK,**[2] **PETR ZHOKHOV,**[3,#] **AND JULIA M. MIKHAILOVA**[2,**]

[1] *IQSE, Department of Physics and Astronomy, Texas A&M University, College Station TX 77843*
[2] *Department of Mechanical and Aerospace Engineering, Princeton University, Princeton, New Jersey, 08544*
[3] *OpenAI, 1455 Third St., San Francisco, CA 94158*
[#] *in personal capacity, unrelated to work at OpenAI*
**zheltikov@physics.tamu.edu*
*** j.mikhailova@princeton.edu*

**We explore the physics behind pulse self-compression to high-power subcycle field waveforms and search for the optimal regimes of such compression over the parameter space as laid out by multimode nonlinear dynamics of ultrashort laser pulses in gas-filled hollow fibers. With the pertinent physical criteria for pulse contrast and no pulse breakup formalized as operational stopping rules, a radical speedup of such search is achieved by delegating its tedious preliminary stages to generative pretrained transformer models. The precise adjustment of self-compression dynamics vis-à-vis the desired compressed pulse output is, however, not yet delegable to AI agents, as the criteria for such search are still largely subjective, operating with fine aspects of subcycle field structure, not yet formalizable as clear search-stopping instructions. An ultimate precision in such subcycle field engineering is attainable via a management of field-phase-sensitive nonlinear processes, enabling pulse self-compression to high-power subcycle field waveforms and opening paths toward fiber-optic attosecond lightwave engineering.**

Subcycle laser pulses, i.e., field waveforms whose envelopes are shorter than their carrier field cycle, provide a powerful resource for ultrafast science and technologies [1–4]. Generation of such pulses at high levels of peak power is notoriously difficult. Cutting-edge laser technologies aimed at achieving this goal combine advanced methods of optical parametric amplification [5–9] with pulse compression in hollow-core fibers (HCFs) [10–15] or laser filaments [16].

As one example relevant to the study presented in this paper, soliton pulse self-compression in a suitably designed ~80-µm-core kagome-cladding hollow-core fiber driven by 1.8-µm, 80-fs OPA output has been shown [10] to enable a robust generation of subcycle pulses with peak powers in the range of a few gigawatts, providing an ultrashort driver for strong-field nonlinear physics. In a related experiment [13], a short-pulse, frequency-tunable idler-wave output of a two-stage OPA was compressed in a 70-µm-core gas-filled antiresonance-guiding HCF to yield carrier-envelope-phase-tunable ultrashort field waveforms of near- to mid-infrared radiation with a peak power up to 4 GW and a pulse width smoothly tunable, via gas-pressure adjustment inside the gas-filled hollow fiber, all the way down to a single field cycle.

Although HCFs with a kagome-lattice [10] or antiresonance [11, 13] cladding are demonstrably efficient in compressing ultrashort laser pulses to subcycle field waveforms at the gigawatt level of peak powers, the nature of light guidance in such fibers does not offer much core-diameter variability, especially for large core diameters, needed for the compression of higher-power laser pulses. In capillary hollow fibers, on the other hand, the core diameter is continuously variable within a broad range, not limited by the design of the cladding.

As a powerful technology of subcycle pulse generation and engineering, subcycle field waveforms with controlled pulse shapes and phase profiles can be synthesized by coherently combing broadband frequency-shifted outputs of a system of gas-filled HCFs running in parallel, thus delivering cycle-tailored ultrafast probes for attosecond spectroscopy [17, 18]. With carefully dispersion- and nonlinearity-managed HCF cascades, millijoule-level picosecond laser pulses can be compressed, as theoretical studies suggest [19], to sub-100 GW subcycle field waveforms.

In this study, we revisit the search for optimal conditions for soliton self-compression to subcycle pulse widths in a single hollow fiber. We explore the physics behind pulse self-compression to high-power subcycle field waveforms and search for the optimal regimes of such compression over the parameter space as laid out by multimode nonlinear dynamics of ultrashort laser pulses in gas-filled HCFs understood in terms of fiber dispersion, full third-order nonlinearity, cycle-resolved photoionization, pulse self-steepening, waveguide loss, nonlinear mode coupling, and soliton decay via dispersive waves.

Aiming to identify the optimal regimes of self-compression of high-power ultrashort laser pulses in gas-filled HCFs, we numerically solve the unidirectional pulse propagation equation (UPPE) [1, 20, 21], which includes the exact dispersion profiles of the pertinent fiber modes, Kerr nonlinearity, pulse self-steepening due to the shock buildup, the $E^3$ third-harmonic term of nonlinear polarization, and field-cycle-resolved photoionization. As a part of our analysis, field-waveform evolution predicted by the UPPE is benchmarked against the predictions of the generalized nonlinear Schrödinger equation (GNLSE) [22, 23], which solves for the pulse envelope and includes exact dispersion of the fiber modes, Kerr nonlinearity, shock-induced pulse self-steepening, and cycle-averaged photoionization.

As a case study, we focus on laser pulses with a central wavelength $\lambda_0 = 2.0$ μm, pulse width $\tau_0 = 50$ fs, pulse energy $U_0 = 300$ μJ, and a variable carrier – envelope phase (CEP) $\phi_0$. Light pulses with such parameters are attainable via advanced fiber-laser, thin-disk-laser, and optical parametric chirped-pulse amplification technologies [15, 24 – 31], finding a broad use in strong-field optical science [10, 13, 26, 29] and x-ray imaging [28, 30, 31]. Self-compression of laser pulses is achieved in the regime of anomalous group velocity dispersion (GVD) in a gas-filled HCF. Dispersion of HCF modes in the mid-IR-to-UV range is a mixture of positive dispersion of the gas filling the fiber core and the negative dispersion of the waveguide. Given the sort of gas (argon in our case study), both the dispersion and nonlinearity of waveguide modes in an HCF are tuned by varying the gas pressure $p$ and hollow-core diameter $d$. In Fig. 1a, the range of anomalous GVD ($\beta_2 < 0$), enabling self-compressing pulse dynamics, is shown by color shading on the $d$–$p$ plane.

Seeking to quantify the soliton power of the input laser pulse within the $d$–$p$ parameter space, we also plot in Fig. 1a the isovalue lines of the ratio $N = \sqrt{l_d/l_{nl}}$, where $l_d$ is the dispersion length and $l_{nl}$ is the nonlinear length. For ideal solitons, defined as solutions to the nonlinear Schrödinger equation (NLSE), $N$ is the key scaling parameter for soliton self-compression, suggesting powerful design rules for fiber-optic pulse compressors [32]. For GNLSE and UPPE soliton dynamics, however, $N$ alone is not sufficient as a predictor of soliton self-compression [33], but is useful as a guide across the available parameter landscape (Fig. 1a).

In the numerical procedure adopted here, the field evolution equations are discretized on a uniform retarded-time grid, with dispersion and attenuation applied as exponential factors in the Fourier space. The ideal NLSE, serving as a universal benchmark, is integrated by an adaptive fourth-order split-step method with step doubling for error control. The GNLSE and the UPPE are integrated in the interaction picture via a fixed-step fourth-order Runge–Kutta scheme. Nonlinear terms in the evolution equations are computed in the time domain and projected onto the fiber eigenmodes. The density of photoelectrons is recomputed at every step of the Runge–Kutta procedure using cycle-averaging and field-phase-sensitive photoionization formalisms in GNLSE and UPPE simulations, respectively. A typical discretization step along the time grid is 0.125 fs for the GNLSE and 0.0625 fs for the UPPE.

To speed up the search over an ample parameter space laid out by self-compression dynamics in HCFs, tedious preliminary stages of this search were delegated to a generative pretrained transformer (GPT) model (GPT-6.1 Astra, GPT-6.1 Sol). To this end, clear operational stopping rules were formulated to guide this search. The most stringent among the stopping rules used in our search requires no zeros for the time derivative of the field envelope except for the sole peak of the pulse, thus guaranteeing no pulse breakup. With this rule applied, the search stops as it reaches the pulse compression regime (orange box in Fig. 1a) achieved with a fiber-core diameter $d \approx 0.22$ mm, fiber length $L \approx 0.62$ m, and argon pressure $p \approx 1.16$ bar, corresponding to $N \approx 7.9$. This regime yields compressed pulses (Fig. 1b) with an FWHM pulse width of $\tau_c \approx 4.8$ fs. With the peak power defined as $P_c = \max_t P(t)$, with the full laser-pulse energy $U = \int_{-\infty}^{\infty} P(t)dt$, we find $P_c \approx 9.8$ GW for the compressed pulse in Fig. 1b, a factor of 1.75 higher than the peak power of input laser pulses. In this regime, however, waveguide and ionization losses limit the total output laser energy to $U_f \approx 116$ μJ, with energy $u_c \approx 38$ μJ within the envelope FWHM.

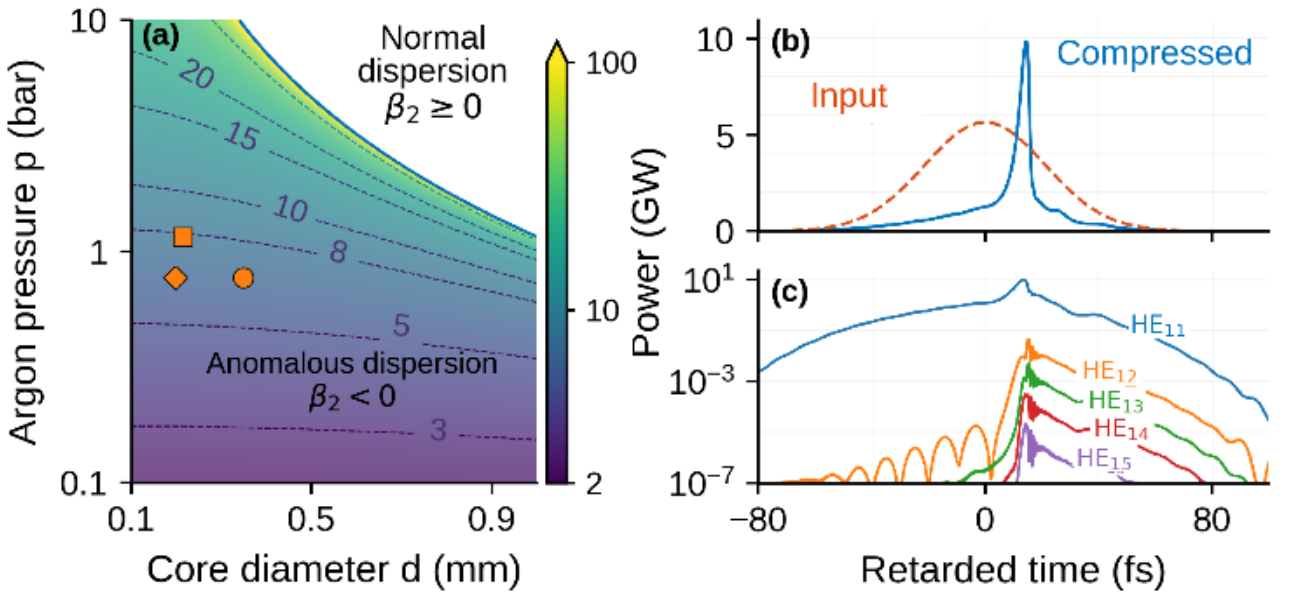

Fig. 1. (a) The fiber-core-diameter – gas-pressure plane view of the parameter space for self-compression of 2.0-μm, 50-fs, 300-μJ laser pulses in a gas-filled HCF. The range of anomalous GVD ($\beta_2 < 0$) is shown by color shading. Color encodes $N = \sqrt{l_d/l_{nl}}$ of the input pulse. Dotted lines are the isovalue lines of $N$. Scatter points show pulse compression optimized for no pulse breakup (box), $Q_t > 0.50$ (circle), and $Q_c > 0.50$ (diamond). (b) Pulse envelope of a laser pulse compressed in an argon-filled fiber with $d \approx 0.22$ mm, $p \approx 1.16$ bar, $L \approx 0.62$ m (box in Fig. 1a). (c) The modal content of the compressed HCF output.

Nonlinear evolution of the field waveform in a gas-filled HCF is coupled to a complex spatial dynamics, in which the spatial profile of the refractive index, induced by Kerr nonlinearity and ultrafast ionization, gives rise to mode coupling, thus opening a channel for energy transfer from the fundamental mode of the HCF to its higher modes. The modal content of the HCF output is nonuniform within the pulse (Fig. 1c), as the higher modes gradually build up along the propagation path and from the leading edge of the pulse to its trailing edge through continuous energy flow from the fundamental mode. Although the modal content of the compressed field waveform becomes complex toward the back of the pulse, the power carried by the fundamental mode is at any time at least two orders of magnitude higher than the total power of all the higher order modes (Fig. 1c).

In search of a higher output energy in a compressed laser pulse, we loosen the search criteria by reformulating the search-stopping rule in terms of a lower bound on the ratio $Q_t$ of the total energy of the compressed output to the input energy $U_0$. With the upper bound on $Q_t$ set at 0.50, the search stops at the pulse-compression regime with $d \approx 0.35$ mm, $L \approx 2.0$ m, $p \approx 0.80$ bar (circle in Fig. 1a), yielding compressed pulses with $\tau_c \approx 2.85$ fs, $u_c \approx 40$ μJ, and $P_c \approx 17.6$ GW within $U_f \approx 153$ μJ output.

To explore field-evolution scenarios yielding deeply subcycle pulse widths, we define the search-stopping rule in terms of a lower bound on the ratio $Q_c$ of the energy $u_c$ within the pulse-envelope FWHM of the compressed pulse to the total energy $U_f$ of the compressed output. The requirement $Q_c > 0.50$ stops the search at $d \approx 0.20$ mm, $L \approx 1.03$ m, $p \approx 0.77$ bar, $N \approx 6.3$ (diamond in Fig. 1a), yielding, as UPPE modeling shows, compressed pulses with $\tau_c \approx 2.1$ fs and $P_c \approx 9.2$ GW for the input CEP $\phi_0 = 0$ and $\tau_c \approx 1.8$ fs and $P_c \approx 10.8$ GW for $\phi_0 = \pi/2$ (Figs. 2a – 2f, 3a – 3d).

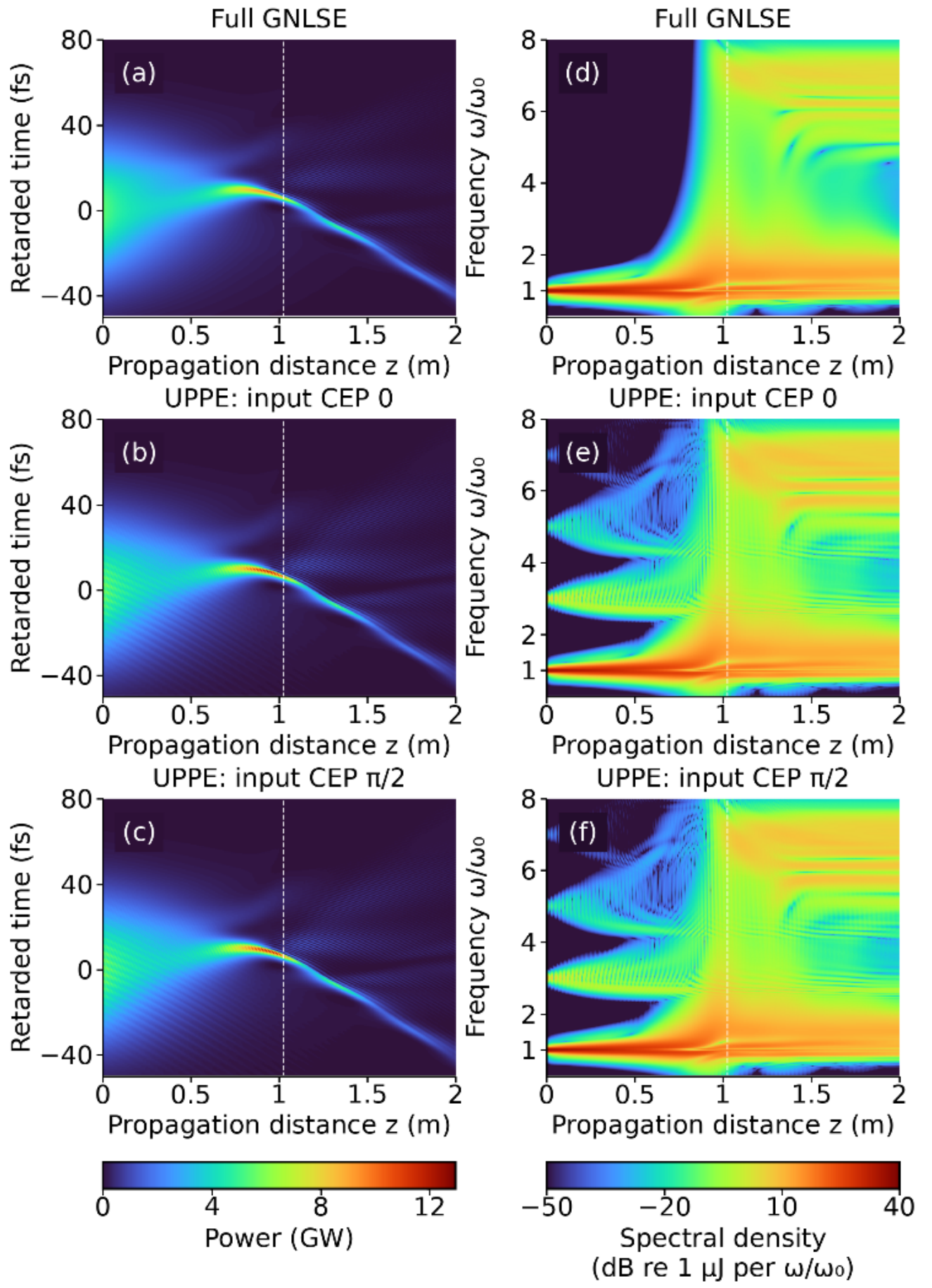

Fig. 2. Temporal (a – c) and spectral (d – f) evolution of a 2.0-μm, 50-fs, 300-μJ laser pulse in a $d \approx 0.20$ mm, $p \approx 0.77$ bar argon-filled hollow fiber as described by the GNLSE (a, d) and the UPPE (b, c, e, f) for $\phi_0 = 0$ (b, e) and $\phi_0 = \pi/2$ (c, f).

To verify the accuracy of these results, we repeated UPPE simulations with refined discretization in time and space, viz., with a propagation increment of 62.5 μm and time step of 0.0625 fs. This decreased compressed-pulse FWHM $\tau_c$ by ≈ 1.7 attoseconds and increased the peak power by less than 0.04%. Moreover, halving the propagation increment and time step simultaneously with extending the UV boundary of the spectral window all the way to 120 nm was found to alter $\tau_c$ by at most 0.1%.

To understand the physics enabling pulse self-compression to such extremely short field waveforms, we examine the temporal (Figs. 2a–2c) and spectral (Figs. 2d–2e) evolution of $\lambda_0 = 2.0$ μm, $\tau_0 = 50$ fs, $U_0 = 300$ μJ laser pulses in a $d \approx 0.20$ mm, $p \approx 0.77$ bar fiber as described by the GNLSE (Figs. 2a, 2d) and the UPPE (Figs. 2b, 2c, 2e, 2f). The initial phase of pulse evolution (within ≈ 10 cm in Figs. 1a–1c) is dominated by soliton pulse self-compression. As a signature of such dynamics, the pulse envelope and the spectrum of the laser-field waveform are largely symmetric. However, as the pulse width approaches the field cycle as a part of this pulse compression dynamics, self-steepening of the laser pulse starts to play a prominent role [34, 35], giving rise to a pronounced asymmetry of pulse envelope (beyond ≈ 10 cm in Figs. 1a–1c) and enhancing the high-frequency part of the spectrum (Figs. 1d–1f), which, in its turn, seeds emission of dispersive waves, readily visible in the short-wavelength part of the spectra in Figs. 1d–1f.

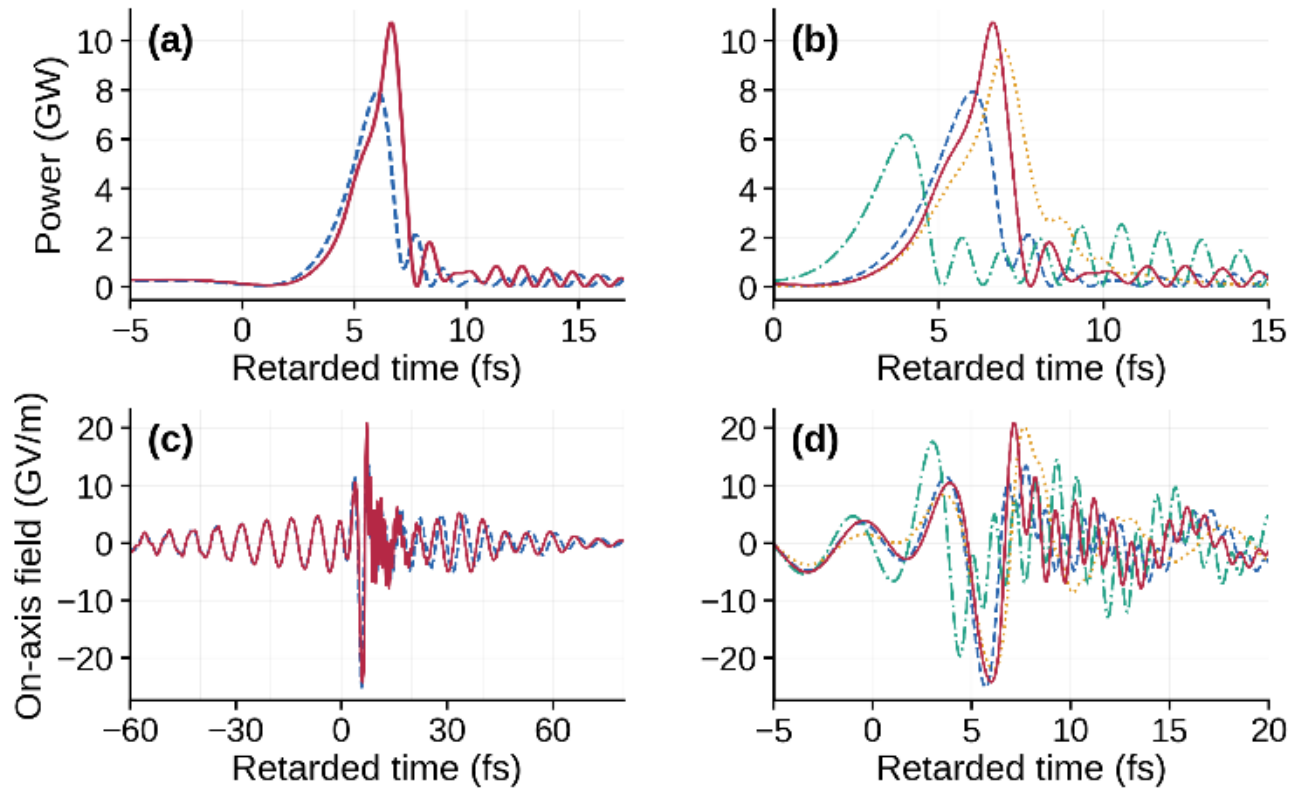

Fig. 3. (a, b) Pulse envelopes and (c, d) field waveforms at the output of a $d \approx 0.20$ mm, $L \approx 1.03$ m, $p \approx 0.77$ bar argon-filled hollow fiber calculated with full UPPE (red solid lines), UPPE-1 (orange dotted lines), UPPE-2 (cyan dash – dotted lines), and UPPE-3 (blue dashed lines) with $\phi_0 = \pi/2$.

In Figs. 3a – 3d, 4a, 4b, we present the pulse envelopes (Figs. 2a, 2b), field waveforms (Figs. 2c, 2d), and spectra (Figs. 4a, 4b) at the output of a $d \approx 0.20$ mm, $L \approx 1.03$ m, $p \approx 0.77$ bar argon-filled hollow fiber (diamond in Fig. 1a) calculated by solving the full UPPE (red solid lines), UPPE in which cycle-resolved photoionization is replaced by cycle-averaged photoionization (UPPE-1, orange dotted lines), UPPE with the $E^3$ third-harmonic term omitted from nonlinear polarization (UPPE-2, cyan dash–dotted lines), and UPPE in which both the $E^3$ term is dropped and cycle-resolved photoionization is replaced by cycle-averaged photoionization (UPPE-3, blue dashed lines).

Cycle-resolving effects in UPPE dynamics are seen to enhance the high-frequency part of the spectrum of the laser pulse (Figs. 2e, 2f, 4a, 4b) and to give rise to odd harmonics of the driver, especially well-resolved at the initial stage of pulse evolution (Figs. 2e, 2f). At later stages of pulse compression, harmonic peaks become broadened due to cross-phase modulation and merge with the shock-wave-enhanced blue wing of the spectrum of the driver, to give rise to an intense mid-IR-to-UV continuum spectrum, which is further enhanced by dispersive waves emitted by IR solitons fissioning from the driver pulse. With all this added spectral content, pulse self-compression dynamics in Figs. 2 – 4 yields even shorter pulse widths. The effects of the $E^3$ term and cycle-resolved photoionization are quantified in Figs. 3b, 3d. Here, calculations with no $E^3$ term and with photoionization cycle-averaged (UPPE-3) give $\tau_c \approx 2.1$ fs and $P_c \approx 9.3$ GW (dotted curve in Fig. 3b). Calculations with both cycle-resolving processes restored, on the other hand, lead to $\tau_c \approx 1.8$ fs and $P_c \approx 10.8$ GW (solid curve in Figs. 3a, 3b).

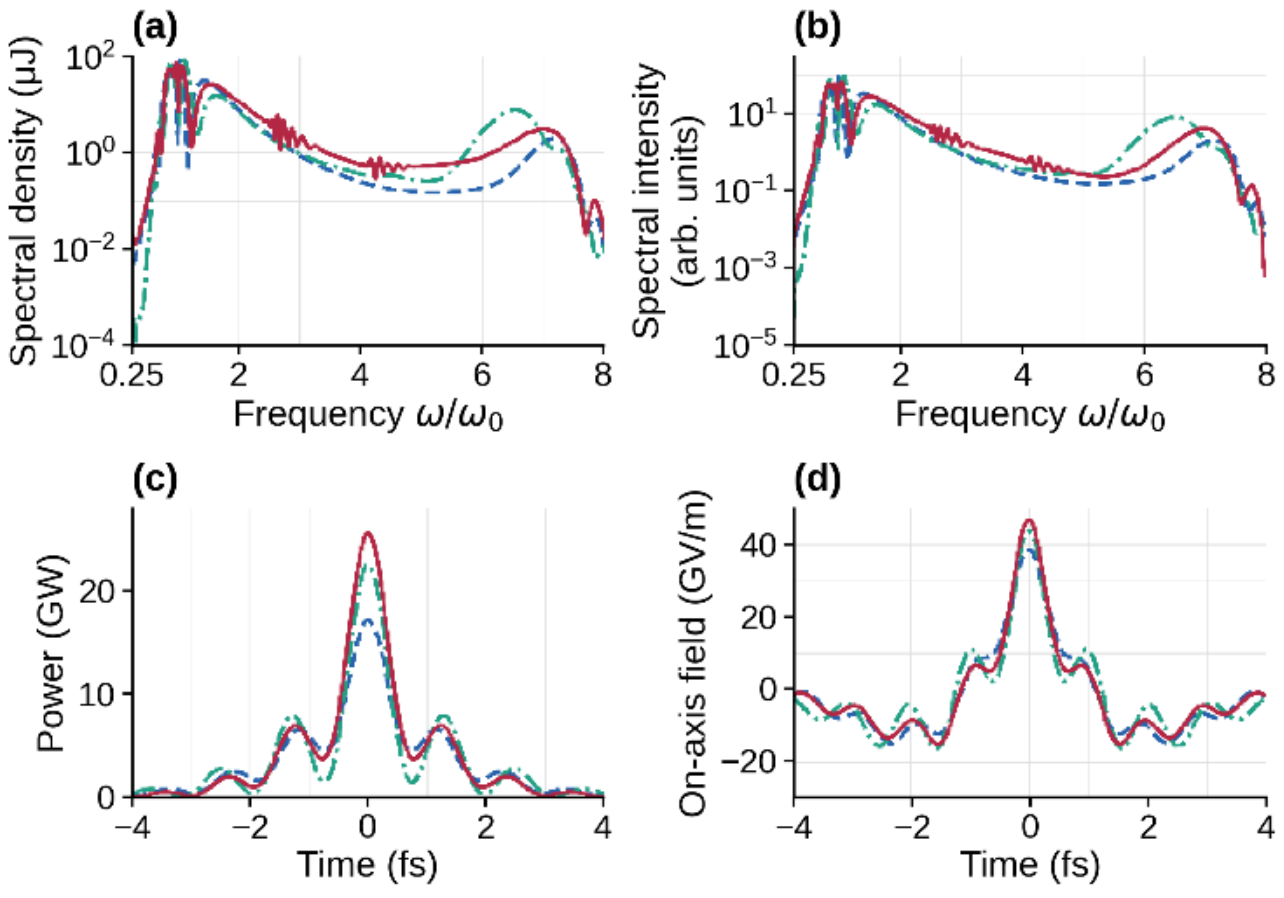

Fig. 4. (a, b) The spectral density of the output of a $d \approx 0.20$ mm, $L \approx 1.03$ m, $p \approx 0.77$ bar argon-filled hollow fiber calculated with full UPPE (red solid lines), UPPE-2 (cyan dash – dotted lines), and UPPE-3 (blue dashed lines) with $\phi_0 = 0$ (a), $\phi_0 = \pi/2$ (b). (c, d) Pulse envelopes (c) and field waveforms (d) for the spectra in Fig. 4b.

The spectral content of the fiber output in Figs. 3a – 3d, 4a, 4b supports, in principle, a synthesis of subfemtosecond field waveforms. As a compelling proof of principle, hollow-fiber-based self-compression to attosecond pulse widths has been demonstrated in recent experiments by Heinzerling et al. [14]. An adequate management of cycle-resolving nonlinear processes, as Figs. 4c, 4d show, is a powerful resource of such high-precision phase correction, opening a path toward high-peak-power cycle-engineered sub-fs optical field waveforms. Indeed, with both cycle-resolved processes disabled in UPPE modeling, the fiber output with a spectrum as shown in Fig. 4b is compressible to a transform-limited FWHM pulse width of ≈1.3 fs with a peak power of ≈17 GW (dashed line in Figs. 4c, 4d). Full UPPE dynamics, on the other hand, with its cycle sensitivity fully restored yields a transform-limited FWHM pulse width of ≈ 0.78 fs and a peak power of ≈ 26 GW.

To summarize, analysis of multimode nonlinear dynamics of ultrashort laser pulses in gas-filled HCFs reveals pulse evolution scenarios whereby high-peak-power ultrashort laser pulses can be compressed to gigawatt subcycle field waveforms. Optimal pulse compression scenarios, yielding high-contrast field waveforms with shortest pulse widths and highest peak powers are shown to unfold via soliton compression coupled to optical shock waves, in which field-cycle-sensitive nonlinear processes shape the field waveform on the subcycle scale, thus holding the keys for precise subcycle field engineering.

**Funding**. DOE Office of Science (DE-SC0024882); Gordon and Betty Moore Foundation (GBMF12255); National Science Foundation (PHY 2512131, PHY 2206711), Department of Energy (DE-SC0025497, DE-SC0025577).

**Disclosures**. The authors declare no conflicts of interest.

**Data Availability Statement**. Data underlying the results of this paper may be obtained from the author upon request.